\documentclass[showpacs,twocolumn,preprintnumbers,showkeys,superscriptaddress,amsmath,amssymb,nofootinbib]{revtex4-2}

\usepackage{color}
\usepackage{amsmath}
\usepackage{amssymb}
\usepackage[utf8]{inputenc}
\usepackage{amsfonts}
\usepackage{bm}
\usepackage{bbold}
\usepackage{hyperref}
\usepackage{graphics}
\usepackage{graphicx}
\usepackage{xspace}

\begin{document}

\title{\Large The transition from non-convex region to hyperbolic region and vice versa in the Beryllium-Graphene-Beryllium three-layer structure: Plasmonics}

\author{Badie Ghavami} \email{badie.ghavami@ipm.ir}
\affiliation{School of Nano Science, Institute for Research in Fundamental Sciences (IPM), Tehran 19395-5531, Iran}

\author{Elnaz Rostampour}
\affiliation{Department of Physics, Urmia University, Urmia, Iran}

\date{\today}

\begin{abstract}
Meta-materials are one of the important discussions of condensed matters which have unusual optical properties.
They can be used to regulate light and wave propagation.
Natural hyperbolic and non-convex plasmons are the characteristics of meta-materials that are observed in these materials. In this paper, we have studied Beryllium-Graphene-Beryllium (BeCBe) three-layer optical properties and have shown this material has hyperbolic plasmon in the THz frequency region and non-convex plasmons. We have shown most of the plasmon energy propagation along the $q_y$ axis for the hyperbolic region. Also, it can be seen that there is an elliptic case only in 1029.71 THz frequency. The starting point of the hyperbolic region occurs at the frequency of 289.31 THz, which is the first and the narrowest hyperbolic window.
Due to the existence of far more free electrons in the BeCBe plasmonic material, the presence of the visible region in the third hyperbolic window causes a high plasma frequency.
\end{abstract}

%\maketitle

\pagestyle{myheadings}

\maketitle

\section{Introduction}
 The plasmon-polariton is an attractive optical phenomenon found in 2D materials. By note in plasmonic properties, we have four classes by attention imaginary conductivity part:
isotropic, anisotropic, hyperbolic, and chiral 
plasmon-polaritons \cite{nemilentsau2019chiral,garcia2014graphene, grigorenko2012graphene,gangaraj2016directive, mojarro2022hyperbolic, rostampour2021effect}.
In these classes, hyperbolic materials interest us, because they enable a wide range of applications that 
include far-field subwavelength imaging, nanolithography,  emission engineering \cite{drachev2013hyperbolic},
negative index waveguides \cite{podolskiy2005strongly}, subdiffraction photonic funnels \cite{govyadinov2006metamaterial}, 
and nanoscale resonators \cite{yao2011three}. Hyperbolic materials show dispersion
and combine properties of transparent reflective metals and 
dielectrics that ellipse materials don't show these properties \cite{drachev2013hyperbolic}.
Therefore, the behavior of hyperbolic materials causes them to be considered. Hyperbolic 
materials are highly anisotropic materials that  are principal components of real parts of their dielectric 
(imaginary part of conductivity) tensor  having opposite signs \cite{poddubny2013hyperbolic, ferrari2015hyperbolic}.
They can naturally support very limited hyperbolic polarization \cite{gjerding2017layered,narimanov2015naturally,dai2014tunable}.
For example, it could be shown for graphene when the imaginary part of the surface conductivity is negative (hyperbolic properties),
it becomes transverse electric surface polaritons in the frequency region \cite{mikhailov2007new}.
Hyperbolic metasurface are different conditions possible, an instance of, it can be realized 
for certain LC-type contours in the microwave frequency region \cite{chshelokova2012hyperbolic,shchelokova2014magnetic}
and formed by anisotropic plasmonic particles 2D lattice in the infrared region (graphene nanoribbons) \cite{trushkov2015two}.
In the terahertz (THz) range, hyperbolic polaritons have observed MgB$_2$,  graphite, 
cuprate, and ruthenate \cite{sun2014indefinite}. Also, in the range of mid-infrared (mid-IR), 
for instance, hyperbolic polaritons were observed in the Van der Waals crystal \cite{zheng2019mid}. 
Provide natural hyperbolic polaritons \cite{tao2019cherenkov} and 
tunable hyperbolic plasmons \cite{dehdast2021tunable} in the THz and mid-IR 
in the 2D materials observed. Black phosphorus (BP) is one of the 2D materials to study 
new plasmonic designs which have tunable optical properties and are  flexible 
\cite{van2019tuning,huang2018black,gomez2016flatland},
because of having high tunable direct bandgap and carrier mobility in-plane anisotropy. Pure phosphorous didn't have hyperbolic plasmons in the near-infrared but they made hyperbolic plasmons with external modulation such as carbon phosphide \cite{dehdast2021tunable}. \\
Despite the hyperbolic behavior of the metamaterial in the hyperbolic metamaterial, the vertical component of the effective permittivity depended on the number of layers of the graphene sheets \cite{gric2017tunable,sreekanth2013negative,zhu2013nanoscale,xiang2014critical}. In a multilayer periodic structure composed of graphene layers in the THz frequency range, the elliptic to hyperbolic regime existed using an external gate voltage \cite{iorsh2013publisher}. In the material examined here, the presence of Be atom makes us not have the elliptic regime, and there is an elliptic case only in frequency 1029.71 THz. Unlike the studied case in three layers of graphene, where there was an elliptical region.\\
In a multi-layer graphene-dielectric composite material, backward waves and forward waves were propagated under the hyperbolic and elliptical dispersion regimes, respectively \cite{othman2013graphene}. The formation of hybrid polaritons in the multilayer structures composed of graphene and hexagonal boron nitride film was more in hyperbolic regions which despite the intrinsic natural hyperbolic behavior of hexagonal boron nitride, there was a more effective hyperbolic behavior \cite{zhao2017near}. In the multilayer structure of graphene and Al$_2$O$_3$, elliptic to hyperbolic dispersion was observed at a wavelength of 4.5 $\mu m$  by an infrared ellipsometer\cite{chang2016realization}.
\\
This paper is written as follows: In section \ref{sec2}, we have briefly explained the theory of dielectric function and optical properties of 2D materials, and in section \ref{sec3} is explained the computational method by details that made the ab-initio method as DFT calculation. Section \ref{sec4} becomes a discussion result of the BeCBe three-layer and investigates plasmonic properties for variable frequency. Finally, the summary and conclusion are provided in section \ref{sec5}.

\section{Computational details}\label{sec2}
We have calculated the dielectric function in reciprocal space 
representation for ground-state electronic structure.
Here, We will briefly explain the theory of calculations. The dielectric function is dependent on wave vector and frequency which is written as 
$\epsilon_{G,G'}(q,\omega)$, where $G$, $q$, and $\omega$ are reciprocal lattice vector, momentum, and frequency of energy 
transfer in excitation energy respectively. 
Therefore, the macroscopic dielectric function is given by $\epsilon_M = [\epsilon_{00}(q,\omega)]^{-1}$ taking into account the local field correction \cite{adler1962quantum}, the absorption spectrum is given by
$\alpha = \Im\epsilon_M(q\to 0, \omega)$, and electron energy loss spectrum (EELS) is gotten by $eels=-\Im [\epsilon^{-1}(q,\omega)]$.\\
The dielectric tensor of materials is:
\begin{equation}
	\epsilon(q,\omega)=\begin{pmatrix}
		\epsilon_{xx}(q,\omega)& 0\\
		0& \epsilon_{yy}(q,\omega)\\
	\end{pmatrix}
\end{equation}
where $\epsilon(q,\omega)= \Re\epsilon(q,\omega)+i\Im\epsilon(q,\omega)$ or the conductivity tensor equal to:
\begin{equation}
	\sigma(q,\omega)=\begin{pmatrix}
		\sigma_{xx}(q,\omega)& 0\\
		0& \sigma_{yy}(q,\omega)\\
	\end{pmatrix}.
\end{equation}
To study surface mode properties, it is considered k-surface, $\omega(q_x,q_y)$=constant, to be shown in Fig.\ref{fig:polar}.
The mode wave-front propagation direction in anisotropic materials is defined by wave-vector $\boldsymbol{q}$ which doesn't generally match with the energy flow direction and it is defined by the group velocity $\boldsymbol{v_{gr}}=\nabla_q\omega(q)$ \cite{nemilentsau2016anisotropic,nemilentsau2019chiral}. \\
The velocity group represents the surface mode in anisotropic materials that transports energy in the orthogonal direction in the k-surfaces.
Thus, the following equation can be used to explain this phenomenon:
\begin{multline}\label{2}
	k_0\sqrt{q_x^2+q_y^2-k_0^2}[\frac{c}{2\pi}-\frac{2\pi}{c}\Im\sigma_{xx}\Im\sigma_{yy}]=\\
	\Im\sigma_{xx}(q_x^2-k_0^2)+\Im\sigma_{yy}(q_y^2-k_0^2)
\end{multline}
where $k_0=\omega/c$ is the free-space wave number ($\omega$ and $c$ are the angular frequency and free-space speed of light respectively).
In the following sections, we have investigated anisotropic plasmons and hyperbolic plasmons which are of the special anisotropic type for BeCBe's three layers.

\section{Computational Method}\label{sec3}
The ab-initio method has done with density functional theory(DFT) calculations which were performed by 
'GPAW' code \cite{enkovaara2010electronic} for 16 bands on the Brillouin zone is sampled by 14 $\times$ 14 $\times$ 1 k-point mesh of the Monkhorst-Pack \cite{monkhorst1976special}.  The cutoff energy is obtained at 500 eV. The exchange-correlation(XC) energy functional 'optPBE-vdW' \cite{dion2004van,wellendorff2012density,klimevs2009chemical} is implemented self-consistently in GPAW considering the Van der Waals potentials.
Also, it has been utilized XC functional with Broyden mixing \cite{perdew1996generalized} as the exchange-correlation energy along with the gride-based projection-augmented wave(PAW) to determine the electronic and optical properties of the BeCBe three-layer structure. 
The structure is optimized by Hellmann-Feynman forces, which are smaller than 0.01 eV/$A^\circ$ on each atom.
The optical properties are calculated by calculating the polarizability response function using the framework of the random phase approximation(RPA). The response function has been evaluated using the Gritsenko-Leeuwen-Lenthe-Baerends-Solis-Correlation potential(GLLB-SC) \cite{gritsenko1995self}
and mesh grid for plasmonic and absorption is utilized  80 $\times$ 70 $\times$ 1 $q$-grid. Finally, we have investigated non-convex and hyperbolic regions for the electrostatically BeCBe three-layer based on Maxwell's boundary condition and the DFT method.

\section{Result and Discussion}\label{sec4}
\begin{figure*}[hbt!]
	\begin{center}
		\includegraphics[scale=0.5]{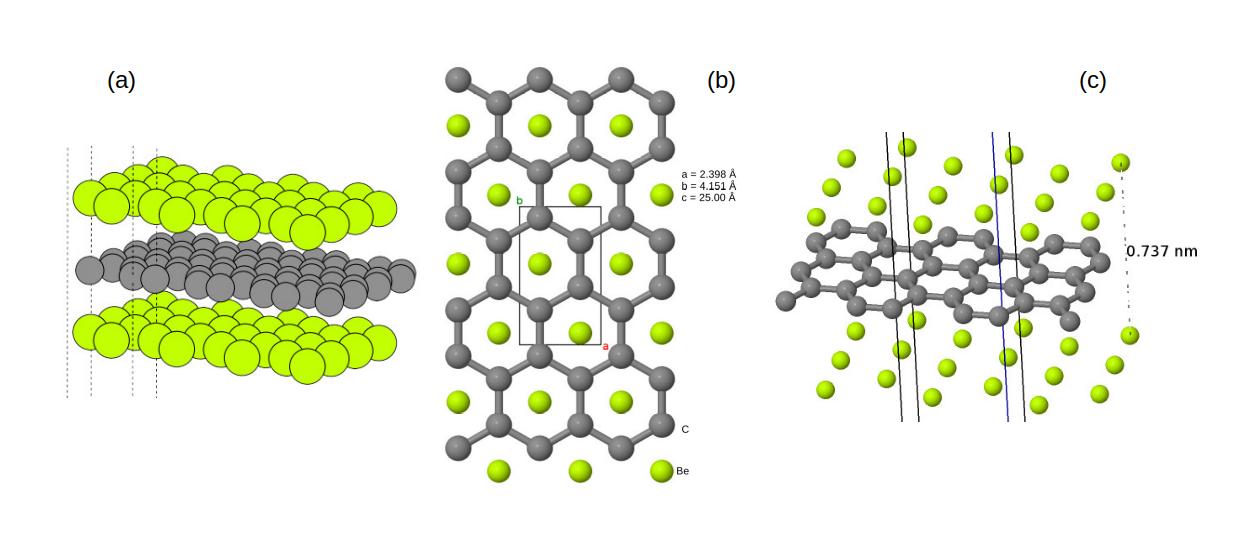}
		%{ \includegraphics[scale=0.5]{BeC_plot}
			%	\includegraphics[scale=0.2]{BeC_plot1}
			%\includegraphics[scale=0.4]{BeC_plot2}}
		\caption{(Color online) It is  shown (a) the geometric of the BeCBe three layer, (b) top view structure, and (c) distance between layers is $3.68 A^\circ$, respectively. Here, Berelium and Carbon atoms are shown with  green and gray colors, respectively. The unit cell parameters are $a = 2.398 A^\circ$, $b = 4.151A^\circ$, $c = 25.0 A^\circ$, and $\alpha=\beta=\gamma=90^\circ$ after optimization. }
		\label{fig:structure}
	\end{center}
\end{figure*}

\begin{figure}[hbt!]
	\centering
	\includegraphics[scale=0.55]{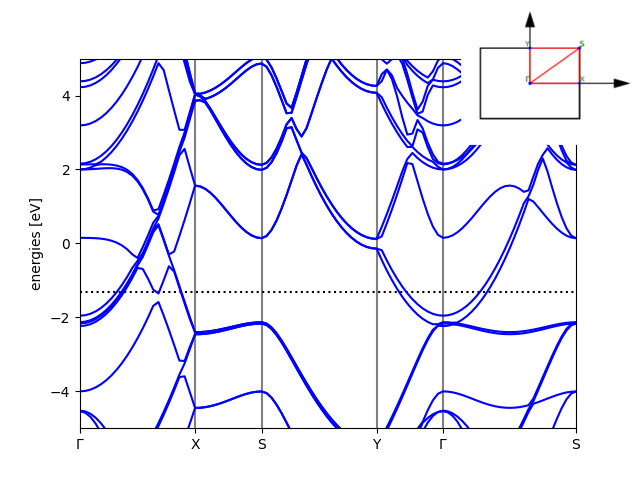}
	\caption{(Color online) Band structure of the BeCBe along high symmetry paths of the Brillouin zone.
		Fermi energy is $E_F= -1.35$ eV. }
	\label{fig:band}
\end{figure}

\textbf{Electronic structure of BeCBe:} In Fig.\ref{fig:structure}, the system under consideration consists of three layers of Beryllium, Graphene, and Beryllium.
There is  3.68 $A^\circ$  between each layer after optimizing energy.  
We have studied the configuration band structure using the DFT method.
Fig.\ref{fig:band} has shown the band structure of the system. 

\begin{figure}
	\centering
	\includegraphics[scale=0.6]{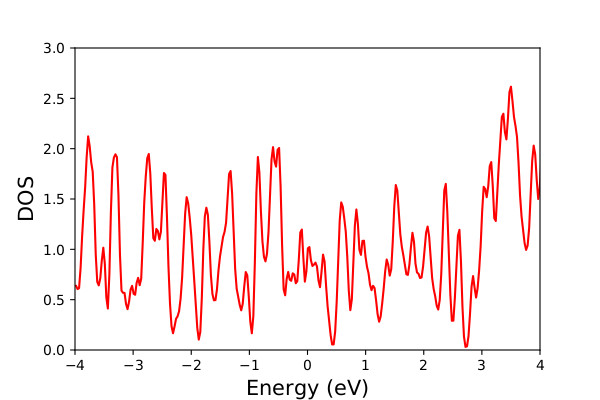}
	\caption{(Color online) Density of states (DOS) diagram BeCBe. It shows to be metal bulk.}
	\label{fig:dos}
\end{figure}

According to the band structure and density of states (DOS)- Fig.\ref{fig:dos}- diagrams,
it is observed that  the gap energy is zero and this structure has metal properties.
Near the valence band maximum (VBM) and the conduction band minimum (CBM) at 
the $\Gamma$ point and X point, we have found the different effective mass values
($m^\star_x$, $m^\star_y$) for the x-direction and  y-direction. We get effective mass through the following equation \cite{WASSERMAN20051} for two directions:
\begin{equation}
	m^\star_{x(y)}=\hbar^2(\frac{\partial^2E}{\partial^2k_{x(y)}})^{-1}
\end{equation}
where fermi energy is  $E_F= -1.35$ eV, and the anisotropy ratio is defined by\cite{frenzel2017anisotropic}:
\begin{equation}
	\frac{\omega_{q,x}}{\omega_{q,y}} = \sqrt{\frac{m^{*}_y}{m^{*}_x}}
\end{equation}   
where $m^*_{x(y)}$ and $\omega_{qx(y)}$ are effective mass and plasmon angular frequency, respectively.
This equation is a good scale to determine plasmonic and polariton-type regimes.
Thus, we would have expected to have plasmon hyperbolic in 
the system.\\
In Fig.\ref{fig:band} exist two linear bands forming Dirac point which are crossed between
$\Gamma$ to $X$  point above the fermi energy level, further $Y$ to $S$ point below the fermi energy level.
The crossing linear bands are both due to the $p_z$ orbitals.
The band gap in paths $\Gamma$ to $X$ point, $Y$ to $\Gamma$ point, and $\Gamma$ to $S$  point
are zero and show material properties in these paths. 
To paths $X$ to $S$ and $S$ to $Y,$ the band structure has gape energy as 
in these paths' system is a semiconductor. 
Also, this view can expect to view plasmonic hyperbolic phenomena that the surface of the system is shown semiconductor and the bulk structure is metal. 
In Fig.\ref{fig:dos}, DOS has shown to be a metal bulk structure. 

\begin{figure}[hbt!]
	\begin{center}
		\includegraphics[scale=0.33]{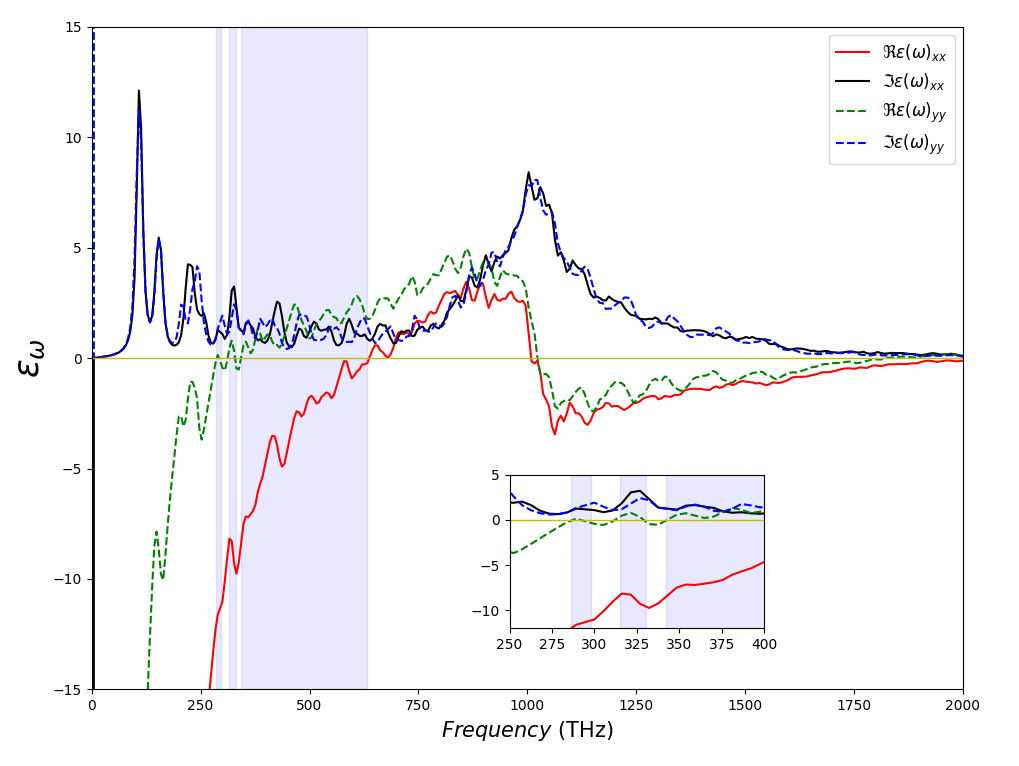}
		\caption{(Color online) Imaginary and real Dielectric function of the  BeCBe three-layer for two directions x, y. The plot shows plasmonic different regions in the THz frequency. }
		\label{fig:abs}
	\end{center}
\end{figure}

\textbf{Optical properties:} We have studied optical properties in our structure. 
For this aim, we calculated and illustrated imaginary and real dielectric functions in two directions x and y. Fig.\ref{fig:abs} have shown the dielectric function of the BeCBe structure.
The polariton behavior in a metamaterial and metasurface depends on the real part of
the dielectric function sign for each direction as for the isotropic plasmonic, the anisotropic plasmonic,  the hyperbolic plasmons, and the chiral plasmons are given by:\\
$\epsilon_{xy}=\epsilon_{yx}=0$, $ \epsilon_{xx}=\epsilon_{yy}$\cite{scholz2013plasmons,wang2015plasmon, garcia2014graphene, hanson2008dyadic}, \\
$\epsilon_{xy}=\epsilon_{yx}=0$, $\Re\epsilon_{xx}\times\Re\epsilon_{yy}>0$\cite{nemilentsau2016anisotropic,low2014plasmons},\\
$\epsilon_{xy}=\epsilon_{yx}=0$, $\Re\epsilon_{xx}\times\Re\epsilon_{yy}<0$\cite{gomez2015hyperbolic},\\
and $\epsilon_{xy}=-\epsilon_{yx} \neq 0$, $\epsilon_{xx}=\epsilon_{yy}$\cite{kumar2016chiral}, respectively. \\

This property can be mentioned based on the tensor of conductive elements:
\begin{equation}
	\epsilon(q_i, \omega)= 1+j\frac{2\pi  q_i \sigma_{ii}(\omega)}{\omega}
\end{equation}
where $i = x ,y$ and $j=\sqrt{-1}$.
\\
We show dielectric function vs frequency in Fig. \ref{fig:abs}. For the free-moving electrons, the polarization response is in opposite direction to the electric field therefore the real part of the dielectric tensor of materials is negative below the plasma frequency. The hyperbolic regime 
is usually the component of the negative dielectric tensor in only one or two spatial directions.

\begin{figure}[hbt!]
	\begin{center}
		\includegraphics[scale=0.49]{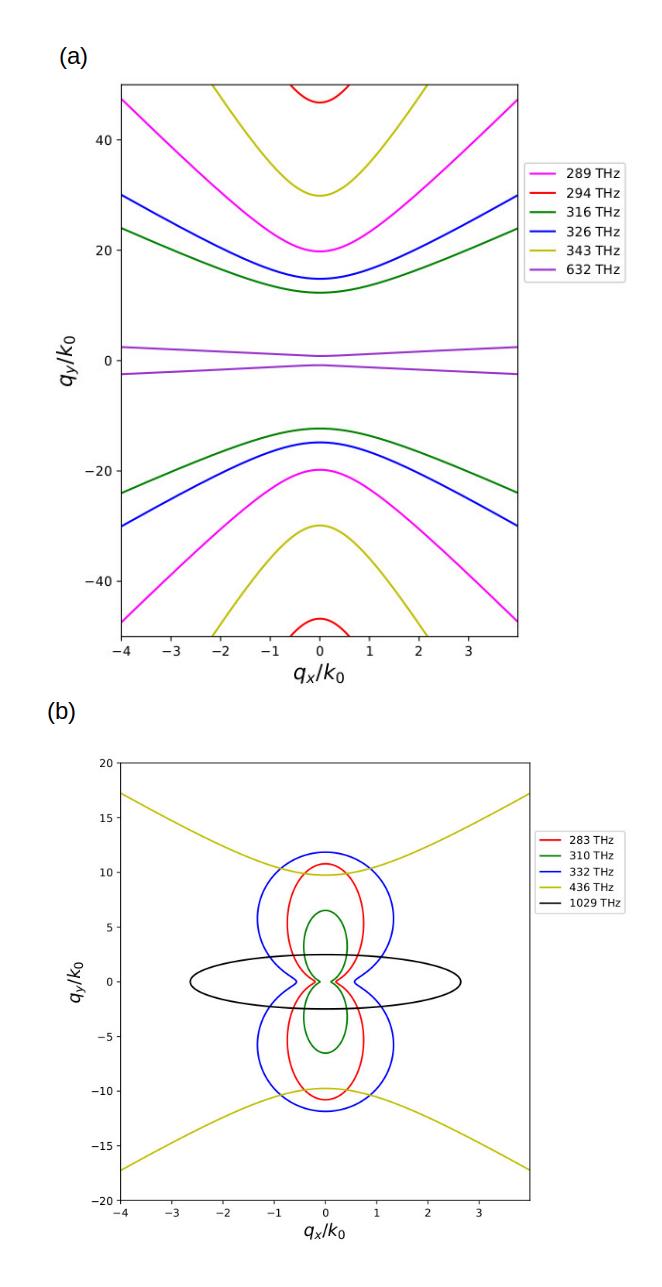}
		\caption{(color online) Diagram of (a) hyperbolic and (b) anisotropic in a few frequencies. When change sign of the dielectric function real part (the conductivity imaginary part), there is an interband transition.} 
		\label{fig:polar}
	\end{center}
\end{figure}
%For 2D- material, the conductivity tensor of a material has followed\cite{kovalev2020proposal}: can be diagonalized in the
%absence of magnetic [34] and nonlocal effects [35].\\
\begin{figure}[hbt!]
	\begin{center}
		\includegraphics[scale=0.37]{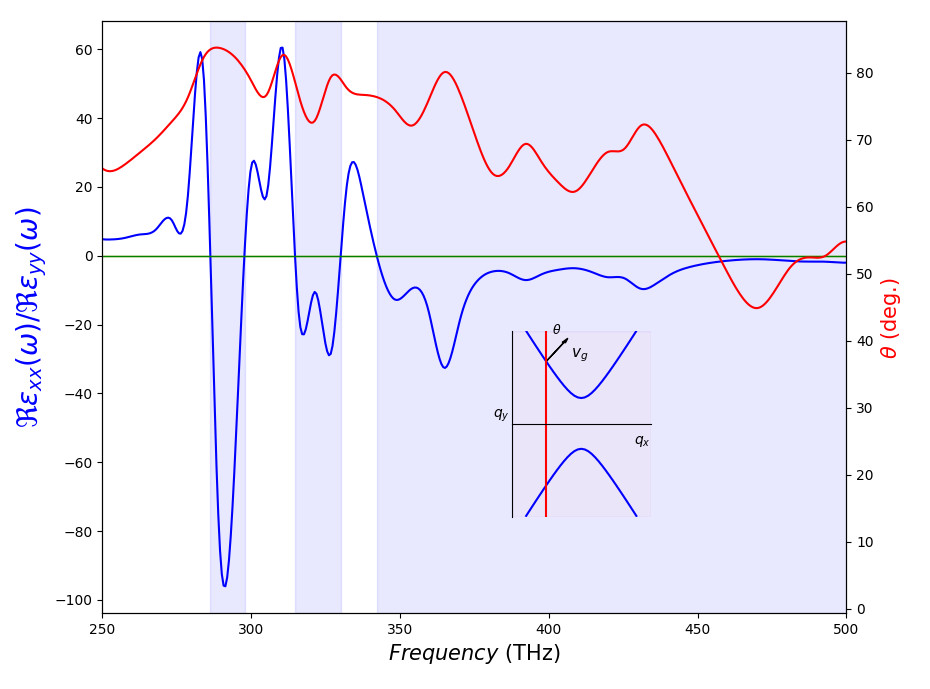}
		\caption{(color online) Selected hyperbolic regime. Graph elongation factor vs frequency (red). Hyperbolas are shown by the angle between the group velocity and the vertical movement axis. }
		\label{fig:rexy}
	\end{center}
\end{figure}
%%%%%%
\textbf{Optical properties of BeCBe:}
%%%%%
The imaginary and real parts of the dielectric function in the $x$ and $y$  directions for BeCBe are drawn in Fig.\ref{fig:abs}, which have different signs in the specified frequency range. 
The obtained results show that BeCBe has three hyperbolic frequency windows at the frequency of $0-2000 $ THz, one of which is wide and the other two are narrow, and $\Re\epsilon_{xx} < 0$, $\Re\epsilon_{yy} >0$  in them.
Hyperbolic frequency windows are shown with blue shadows. The first hyperbolic window is at the frequency of $289.31-294.64$ THz, the second hyperbolic window is at the frequency of $316.06-326.83$ THz, and the third hyperbolic window is at the frequency of $343.05-632.47$ THz.
The behavior of the real parts of the dielectric function is as follows: The value of $\Re\epsilon_{xx}$  at frequencies less than $632.47$ THz is always negative, while the value of $\Re\epsilon_{yy}$ at frequencies less than  $289.31$ THz is negative and is placed in the non-convex region.\\
With the increase of frequency, the value of $\Re\epsilon_{yy}$ is positive and leads to the first hyperbolic region, the value of which reaches zero at the frequency of $294.64$ THz, and with the increase of frequency, it becomes negative and then reaches zero. In this case, the non-convex region is formed. At the frequencies of $316.06$ THz to $326.83$ THz, the value of $\Re\epsilon_{yy}$ is positive and the second hyperbolic region is formed, with the increase of the frequency, the value of $\Re\epsilon_{yy}$ becomes negative and then reaches zero at the frequency of $343.05$ THz, which forms the non-convex region. Again, with the increase of frequency, the value of $\Re\epsilon_{yy}$ is positive up to the frequency of $1023$ THz, and the value of $\Re\epsilon_{xx}$ becomes positive from the frequency of $632.47$ THz onwards and forms the hyperbolic region up to the frequency of $632.47$ THz.
In fact, when $\Re\epsilon_{xx}$  and $\Re\epsilon_{yy}$ have the same sign, an ellipse region is formed, and when the sign of $\Re\epsilon_{yy}$ changes, a hyperbolic region is formed. The non-convex regions have an almost eight-like shape.  Due to the fact that inductive and capacitive behaviors occur at the frequencies between the transition points - the hyperbolic frequencies-, in this material, in the region where $\Re\epsilon_{xx}$ is negative, we have the inductive response and in the region where $\Re\epsilon_{yy}$ is positive, we have the capacitive response. The change in the sign of the dielectric function in this material indicates the change in the behavior of polaritons. For BeCBe, the first hyperbolic window and the second hyperbolic window are in the infrared region. The third hyperbolic window shows the transition from the infrared region to the visible region. In this material,   $\Re\epsilon_{xx}$  reaches zero later than  $\Re\epsilon_{yy}$.
The monolayer of metal-free two-dimensional carbon phosphide has one hyperbolic frequency window and this window is wider compared to the third BeCBe hyperbolic frequency window \cite{dehdast2021tunable}. On the other hand, the width of the third hyperbolic frequency window of BeCBe is more than $MgB_2$ \cite{gao2019tunable}, thin films of $WTe_2$ \cite{wang2020van}, and monolayer black phosphorus  \cite{wang2021prediction}. So, the order of the hyperbolic frequency window of these five materials can be written as 
Monolayer carbon phosphide $>$ BeCBe $>$  $MgB_2$ $>$ monolayer black phosphorus $>$ thin films of $WTe_2$.\\
Fig.\ref{fig:polar} shows isofrequency counters of surface plasmon polaritons. BeCBe has frequencies as   $\omega_{h1}=289.31$ THz, $\omega_{h2}=283.999$ THz, $\omega_{h3}=305.33$ THz, $\omega_{h4}=316.06$ THz, $\omega_{h5}=322.22$ THz,  and $\omega_{h6}=343.05$ THz  in the non-convex region. These frequencies are vertical non-convex contours.
In the hyperbolic region, frequency $\omega_{h\nu p}=436.60$ THz is in hyperbolic form with large openings in Fig.\ref{fig:polar}.
In fig.\ref{fig:polar}, these frequencies are in hyperbolic form with large openings. In this metamaterial, with the change of the sign of the real part of the dielectric function, there are interband transitions. 
The transition frequencies are as Table \ref{tab1}.
\begin{table}[h]
	\small
	\caption{\ Transmission frequencies (THz) are shown for the hyperbolic plasmonic shape with a large opening.}
	\label{tab1}
	\begin{tabular*}{0.48\textwidth}{@{\extracolsep{\fill}}llllll}
		\hline
		$\omega_{tt1} $ &$\omega_{tt2}$ & $\omega_{tt3}$&$\omega_{tt4}$ & $\omega_{tt5}$& $\omega_{tt6}$\\
		\hline
		289.31 & 294.64 & 316.06 & 326.83 & 343.05& 632.46\\
		\hline
	\end{tabular*}
\end{table}

In fig.\ref{fig:polar}, these frequencies are in hyperbolic form with small openings. In case of transition between interband and intraband, hyperbolic dispersion occurs.

\begin{figure*}[hbt!]
	\begin{center}
		\includegraphics[scale=0.55]{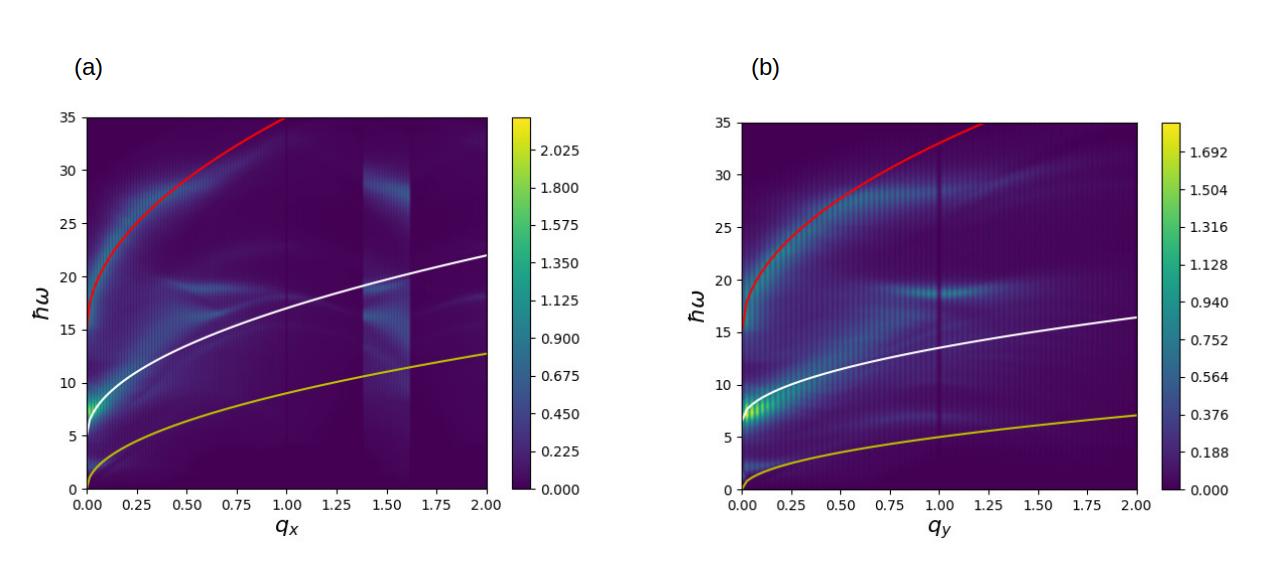}
		\caption{(color online) The plasmon dispersion for the $q_x$ and $q_y$ directions. Plasmon dispersion is more in the $y$ direction than in the $x$ direction.  Plasmons dispersion in the $x$ direction to $q_x=1 A^{\circ -1}$ and in the $y$ direction to $q_y=1.5 A^{\circ -1}$ and then enter the single-particle excitations region. Here, it is seen $\omega \propto\sqrt{q}$.}
		\label{fig:plasmon}
	\end{center}
\end{figure*}

In fig.\ref{fig:polar}, with the increase in frequency, the curves change from non-convex to hyperbolic, from hyperbolic to non-convex, and so on.
The degree of elongation of the isofrequency contours is expressed by the elongation factor. The elongation factor determines the non-convex and hyperbolic regions. At frequencies where the elongation factor is positive, we have the non-convex regime, and at frequencies where the elongation factor is negative, we have the hyperbolic regime.  In Fig. 6, the elongation factor is plotted in terms of frequency.  According to the obtained results, the elongation factor of this metamaterial has the maximum value in three vertical frequencies as $\omega_{\nu1}=282.9$ THz, $\omega_{\nu2}=310.5$ THz, and $\omega_{\nu3}=334.5$ THz. These frequencies have a dumbbell shape as shown in Fig.\ref{fig:polar}.
Before this frequency and after this frequency, frequencies are placed in the non-convex regime.
%Frequencies before this frequency and after this frequency that are placed in the %non-convex regime. 
At frequencies around $\omega_\nu$ ($\Re\epsilon_{yy} \gg \Re\epsilon_{xx}$), the phenomenon of low-loss canalization occurs along the $q_y$ direction. \\
In the transition frequencies $\omega_{tti}$ for $i=1..6$ ($\Re\epsilon_{xx}<0$, $\Re\epsilon_{yy} \approx 0$), we have the phenomenon of canalization and most of the limited plasmon energy is propagated along the  $q_y$ axis. 
In this case, the direction of group velocity and plasmon energy propagation coincide. This situation is more in hyperbolic isofrequency than in non-convex isofrequency. The propagation of surface plasmon polaritons is between the asymptotes of the hyperbola $q_y = \pm q_x tan(\theta)$.
Hyperbolas are expressed by the angle between the velocity of the group  $v_g$ and the axis of the vertical movement $q_y$ ($\theta=tan^{-1}\sqrt{\frac{\Re\epsilon_{xx}}{\Re\epsilon_{yy}}}$) as shown in Fig.\ref{fig:rexy}. Non-convex and hyperbolic surface plasmon polaritons are expressed by equation \ref{2}.
In the range of $0-632.47$ THz, $\Re\epsilon_{xx}$ and $\Re\epsilon_{yy}$  do not cut each other ($\Re\epsilon_{xx}\neq \Re\epsilon_{yy}$)  and there is no circular crossing frequency.  So the elongation factor near one is in the form of vertical lines. \\
According to the obtained results, the behavior of polaritons in BeCBe is anisotropic plasmonic in non-convex regions and hyperbolic plasmons in hyperbolic regions. The hyperbolic regime indicates directional plasmons. 
In Fig.\ref{fig:plasmon}, plasmons propagate in the $x$ and $y$ directions. Plasmon dispersion is more in the $y$ direction than in the $x$ direction. $\Re\epsilon_{yy}$ is equal to zero at different frequencies. Plasmons propagate in the $x$ direction to $q_x=1 A^{\circ -1}$ and in the $y$ direction to $q_y=1.5 A^{\circ -1}$ and then enter the single-particle excitations region. In BeCBe, small acoustic plasmons with very low intensity exist in both the $q_x$ and $q_y$ directions at very low energies. BeCBe exhibits strong anisotropic plasmon. \\
As we expect, at low $q_x$ and $q_y$, with the proximity of the surface plasmon polariton to the  Sommerfeld-Zenneck wave \cite{schumacher1998surface}, it has a photon-like behavior, and with the increase of $q_x$ and $q_y$ and the bending of the dispersion relation curve, it approaches the asymptotic limit of the surface plasma frequency.

\section{Conclusions}\label{sec5}
In summary, hyperbolic and non-convex plasmonic properties are revealed in this material so that the results show three hyperbolic frequency windows. The widest hyperbolic frequency window is located in the frequency range of $343.05-632.47$ THz. We investigate the behavior of polaritons in non-convex and hyperbolic regions with a two-dimensional description of the dielectric function. The opposite sign of the real parts of the dielectric function in both $x$ and $y$ directions indicates hyperbolic plasmon behavior and their same sign indicates anisotopic plasmonic behavior. Plasmons are dispersed in both $x$ and $y$  directions for this hyperbolic metamaterial. In the hyperbolic region, most of the plasmon energy propagates along the $q_y$  axis. In the non-convex regions around the frequencies $\omega_{\nu1}$, $\omega_{\nu2}$, and $\omega_{\nu3}$  the phenomenon of low-loss canalization in the $q_y$ direction is created. The presence of non-Euclidean geometry in BeCBe makes it more practical.
%\section*{Acknowledgments}

\bibliography{ref}

\end{document}